\documentclass[twocolumn,showpacs,preprintnumbers,amsmath,amssymb]{revtex4-1}
\usepackage{graphicx}
\usepackage{bm}
\usepackage{latexsym}
\usepackage{epsfig}
\usepackage{xcolor} 
\usepackage{changes}

\begin{document}

\title{Gravitational waves in $f(R,T)$ {\bf and $f(R,T^\phi)$ theories} of gravity}

\author{M.E.S. Alves}\email{marcio.alves@ict.unesp.br}
\affiliation{Instituto de Ci\^encia e Tecnologia, UNESP - Univ Estadual Paulista, S\~ao Jos\'e dos Campos, SP, 12247-016, Brazil}

\author{P.H.R.S. Moraes}\email{moraes.phrs@gmail.com}
\affiliation{ITA - Instituto Tecnol\'ogico de Aeron\'autica - Departamento de F\'isica, S\~ao Jos\'e dos Campos, SP, Brazil}

\author{J.C.N. de Araujo}\email{jcarlos.dearaujo@inpe.br}
\affiliation{INPE - Instituto Nacional de Pesquisas Espaciais - Divis\~ao de Astrof\'isica, S\~ao Jos\'e dos Campos, SP, Brazil}

\author{M. Malheiro}\email{malheiro@ita.br}
\affiliation{ITA - Instituto Tecnol\'ogico de Aeron\'autica - Departamento de F\'isica, S\~ao Jos\'e dos Campos, SP, Brazil }

\date{\today}

\begin{abstract}
\baselineskip0.4cm
\noindent There is a host of alternative theories of gravitation in the literature, among them the $f(R,T)$ and $f(R,T^\phi)$ theories recently elaborated by T. Harko and collaborators. In these theories, $R$, $T$ and $T^\phi$ are respectively the Ricci scalar, and the traces of the energy-momentum tensors of matter and of a scalar field. There is already in the literature a series of studies of different forms of the $f(R,T)$ and $f(R,T^\phi)$ functions as well as their cosmological consequences. However, there are no studies so far related to gravitational waves. Here we consider such an issue, in particular, studying the putative extra polarization modes that can well appear in the scope of such theories. Different functional forms of $f(R,T^\phi)$ are considered and the gravitational waveforms are found for the extra polarization modes in the cases they are present.
\end{abstract}

\maketitle

\section{Introduction}

Recently, gravitational waves (GWs) were directly detected for the first time \cite{abbott/2016}. Forthcoming observations will contribute to the study and understanding of a large number of research fields in Physics, Astrophysics and Cosmology, from the absolute ground state of matter \cite{mm/2014} to the upper limits on the brane tension values of braneworld cosmologies \cite{mm/2014b}. In the near future, we may also be able to estimate parameters of compact binary systems \cite{berry/2015}-\cite{poisson/1995}, constrain the equation of state of neutron stars \cite{takami/2014}-\cite{read/2009} and distinguish General Relativity (GR) from alternative theories of gravity \cite{alves/2009}-\cite{yagi/2010}.

It is well known that alternative theories of gravity arise as possibilities for evading some standard cosmology shortcomings \cite{clifton/2012}-\cite{arkani-hamed/2000}. Recently elaborated by T. Harko and collaborators, the $f(R,T)$ gravity \cite{harko/2011} is one of the promising alternatives.

Although plenty of well behaved cosmological models have been derived from such a theory (see \cite{ms/2016}-\cite{shabani/2014} and references therein), no efforts have been made in applying $f(R,T)$ gravity to the study of GWs.

It is the purpose of the present article to explore the physical features of GWs in $f(R,T)$ gravity and in different possible formulations of $f(R,T^\phi)$ gravity. We will show that the physics of GWs is strongly dependent on the functional forms of $f(R,T)$ and $f(R,T^\phi)$, in such a way that different formulations can exhibit different numbers of polarization states. In order to characterize the polarization states of GWs for some formulations of interest, we will evaluate the Newman-Penrose (NP) quantities \cite{eardley/1973}-\cite{newman/1962b} predicted by them. For now, it is worth mentioning that the NP formalism has been applied to different alternative theories of gravity, leading to interesting and testable results \cite{alves/2009,alves/2010,depaula/2004}.

\section{$f(R,T)$ gravity}
\label{sec:frt}

Proposed by T. Harko and collaborators \cite{harko/2011} as a generalization of the $f(R)$ theories (see \cite{nojiri/2009,nojiri/2007} and references therein), the gravitational part of the $f(R,T)$ action depends not only on a generic function of the Ricci scalar $R$, as in $f(R)$ gravity theories, but also on a function of $T$, the trace of the energy-momentum tensor $T_{\mu\nu}$. According to the authors, the dependence on $T$ arises from the consideration of quantum effects (conformal anomaly) which are usually neglected in $f(R)$ or GR theories, for instance. The full action in $f(R,T)$ gravity reads
\begin{equation}\label{frt1}
S=\int d^{4}x\sqrt{-g} \left[ f(R,T) + {L}_{m} \right],
\end{equation}
with $f(R,T)$ being an arbitrary function of $R$ and $T$, $g$ is the determinant of the metric $g_{\mu\nu}$, with Greek indices assuming the values $0-3$ and $\sqrt{-g}L_m$ is the Lagrangian density of matter. Throughout this work we will use units such that $4\pi G=c=1$. 

By varying Eq.(\ref{frt1}) with respect to the metric, one obtains the $f(R,T)$ field equations
\begin{align}\label{frt2}
&&f_R(R,T)R_{\mu\nu} - \frac{1}{2}f(R,T)g_{\mu\nu}+\nonumber\\
&&+(g_{\mu\nu}\Box-\nabla_\mu\nabla_\nu)f_R(R,T)=\nonumber\\
&&-\frac{1}{2}T_{\mu\nu}-f_T(R,T)T_{\mu\nu}-f_T(R,T)\Theta_{\mu\nu},
\end{align}
with $\Box\equiv\partial_\mu(\sqrt{-g}g^{\mu\nu}\partial_\nu)/\sqrt{-g}$, $\Theta_{\mu\nu}\equiv g^{\alpha\beta}\delta T_{\alpha\beta}/\delta g^{\mu\nu}$, $f_R(R,T)\equiv\partial f(R,T)/\partial R$, $f_T(R,T)\equiv\partial f(R,T)/\partial T$,  $R_{\mu\nu}$ is the Ricci tensor, $\nabla_\mu$ is the covariant derivative with respect to the symmetric connection associated to $g_{\mu\nu}$ and the energy-momentum tensor, as usual, reads
\begin{equation}\label{frtx1}
T_{\mu\nu} = \frac{2}{\sqrt{-g}} \frac{\delta\left(\sqrt{-g}L_m\right)}{\delta g^{\mu\nu}}.
\end{equation}

For reasons that will be presented below, in this article we are also concerned with a different theory for which Harko et al. \cite{harko/2011} have considered the coupling of gravity with a self-interacting scalar field $\phi$, namely $f(R,T^{\phi})$ gravity, with $T^{\phi}$ standing for the trace of the energy-momentum tensor of the scalar field.

Such a formulation was developed in a cosmological perspective in \cite{ms/2016} and gave rise to a complete scenario of the Universe evolution, able to describe the inflationary, radiation, matter and dark energy eras.

For this case, the full action is given by
\begin{equation}
S=\int d^{4}x\sqrt{-g} \left[ f(R,T^\phi) + L(\phi,\nabla_\mu \phi) +  {L}_{m} \right],
\end{equation}
where $L(\phi,\nabla_\mu \phi)$ is the usual Lagrangian for the scalar field, namely
\begin{equation}\label{frt4}
{L}(\phi,\nabla_\mu \phi)=\frac{1}{2}\nabla_\alpha\phi\nabla^{\alpha}\phi-V(\phi),
\end{equation}
and $V(\phi)$ is a self-interacting potential. Notice that in this theory, the matter fields have only a minimal coupling with gravity and they do not couple with $\phi$.

From (\ref{frtx1}) and (\ref{frt4}), the energy-momentum tensor for the scalar field reads
\begin{equation}
T^{\phi}_{\mu\nu} =\nabla_\mu\phi\nabla_\nu\phi - \frac{1}{2}g_{\mu\nu}\nabla_\alpha\phi\nabla^\alpha\phi + g_{\mu\nu}V(\phi)
\end{equation}
and its trace is given by
\begin{equation}\label{frt6}
T^{\phi}=-\nabla_\alpha\phi\nabla^{\alpha}\phi+4V(\phi).
\end{equation}

\section{GWs in the $f(R, T)$ gravity}

In order to find the number of polarization modes of GWs of a theory we need to examine the far-field, linearized, vacuum field equations of the theory. For vacuum, the $f(R,T)$ and $f(R)$ field equations \cite{nojiri/2009,nojiri/2007} are the same, namely:
\begin{equation}
f_R R_{\mu\nu} - \frac{1}{2}f g_{\mu\nu} - \nabla_\mu \nabla_\nu f_R + g_{\mu\nu}\Box f_R = 0.
\end{equation}

The calculation of the NP parameters for such a theory was carried out in Ref.\cite{alves/2009}. The authors considered $f(R) = R - \alpha R^{-\beta}$, with $\alpha$ and $\beta$ being constants, and found out that if $\alpha \neq 0 $ and $\beta \neq 0$, one has for the NP quantities,
\begin{equation}\label{9}
\Psi_2 \neq 0,~~~\Psi_3 = 0,~~~\Psi_4 \neq 0,~{\rm and}~\Phi_{22} \neq 0,
\end{equation} 
showing that this theory presents the scalar longitudinal mode ($\Psi_2$) and the ``breathing'' scalar transversal mode ($\Phi_{22}$) in addition to the the usual tensor modes represented by $\Psi_4$. But it is worth emphasizing that since $\Psi_2 \neq 0$, this is the only observer-independent mode. The presence or absence of all other modes depends on the observer (see, e.g., \cite{alves/2009,eardley/1973}). 

Therefore, the standard $f(R,T)$ formalism does not give new information about GW polarization states, because in vacuum the $f(R)$ formalism is retrieved.

\section{GWs in the $f(R,T^\phi) = -R/4 + f(T^\phi)$ theory}

Because one expects scalar field terms to appear in the $f(R,T^{\phi})$ field equations for vacuum, new polarization states of GWs can be present in such theories. The present and the following sections will deal with this issue.

In what follows we consider GWs in the absence of matter and therefore we take $L_m =0$. The field equations of the $f(R, T^\phi) = -R/4 + f(T^\phi)$ theory reads \cite{ms/2016}:
\begin{equation}\label{frt7}
G_{\mu\nu}=2[T^\phi_{\mu\nu}-g_{\mu\nu}f(T^{\phi}) - 2f_T(T^{\phi})\nabla_\mu\phi\nabla_\nu\phi],
\end{equation}
where $G_{\mu\nu}$ is the usual Einstein tensor. It is also useful to know explicitly the Ricci scalar, namely
\begin{equation}\label{field eq Ricci}
R = -2[T^\phi - 4f(T^\phi) - 2f_T(T^\phi)\nabla_\alpha\phi\nabla^\alpha\phi].
\end{equation} 

The equation of motion for the scalar field can be found by applying the covariant divergence of the field equations (\ref{frt7}). One then obtains 
\begin{equation}\label{General Field Eq for phi}
(1 - 2f_T)\Box\phi + (1 - 4f_T) \frac{\partial V}{\partial \phi} - 2f_{TT}\nabla^\nu \phi \nabla_\nu T = 0,
\end{equation}
with $f_{TT}\equiv\partial^2 f(R,T)/\partial T^2$.

The properties of GWs depend upon the particular choice for $f(T^\phi)$, for which we analyse some well motivated possibilities below. 

\subsection{The $f(T^\phi) = 2\lambda T^\phi$ case}

One interesting particular case is to consider that $f$ depends linearly on the trace of the energy-momentum tensor of the scalar field, i.e.,  $f(T^\phi) = 2\lambda T^\phi$, where $\lambda$ is a constant. Such a functional form has already been used in $f(R,T^\phi)$ models, yielding a description of a complete cosmological scenario, as can be seen in \cite{ms/2016}. For this case, Eq.(\ref{General Field Eq for phi}) reduces to:
\begin{equation}\label{phi eq motion}
\Box \phi + \left(\frac{1-8\lambda}{1-4\lambda}\right)\frac{\partial V(\phi)}{\partial \phi} = 0,
\end{equation}
for $\lambda \neq 1/4$.

Since we need to analyse the field equations in the linearized regime, we expand the potential around its minimum, obtaining \cite{eardley/1973}
\begin{equation}
\Box \phi + \left(\frac{1-8\lambda}{1-4\lambda}\right)m^2 (\phi-\phi_0) = 0,
\end{equation}
where now $\Box = \eta^{\mu\nu} \partial_\mu \partial_\nu$ with $\eta^{\mu\nu}$ being the Minkowski metric, $m^2 = (\partial^2 V/\partial \phi^2)_{\phi = \phi_0}$ and $\phi_0$ locates the minimum of the potential, which could be obtained from cosmological boundary conditions. A solution of this equation reads
\begin{equation}\label{model of solution for phi}
\phi(x) = \phi_0 + \phi_1 e^{iq_\alpha x^\alpha},
\end{equation}
where $\phi_1$ is a small amplitude and $q_\alpha$ is the wave vector which obeys the following equation 
\begin{equation}
q_\alpha q^\alpha = \left(\frac{1 - 8\lambda}{1-4\lambda}\right)m^2,
\end{equation}
with $\lambda \leq 1/8$ or $\lambda > 1/4$.

The Ricci scalar (\ref{field eq Ricci}) for $f(T^\phi) = 2\lambda T^\phi$ reads
\begin{equation}
R = 2\left[(1-4\lambda) \nabla_\alpha \phi\nabla^\alpha \phi - 4(1-8\lambda)V(\phi) \right], 
\end{equation}
and by considering terms up to first order in $\phi$ we find a constant curvature scalar, namely
\begin{equation}
R = -8(1-8\lambda)V_0 + \mathcal{O} (\phi_1^2),
\end{equation}
where $V_0$ is the minimum value of the potential.

Therefore, from the point of view of the propagation of GWs, the overall effect of the inclusion of a minimally coupled scalar field to first order is equivalent to consider an effective cosmological constant (CC)
\begin{equation}
\Lambda =  2(1-8\lambda)V_0,
\end{equation}
which is positive for $\lambda < 1/8$, null for $\lambda = 1/8$ and negative otherwise. As it is  well known, $\Lambda$ does not introduce any additional polarization states for GWs \cite{Naf2009}. Therefore, we can conclude that GWs in the  $f(R,T^\phi)=-R/4 + 2\lambda T^\phi$ theory have only the two usual polarizations of GR, i.e., $+$ and $\times$.

The above result shows that the linearized field equations of this theory, in the absence of matter, have the Minkowski metric as background only if $\lambda = 1/8$ or $V_0 = 0$. In this case, the GW equations are exactly the same as those of GR theory. 

Otherwise, in order to obtain the first order equations for the GWs, we need to expand the metric around the de Sitter metric. Here we do not consider such an issue, instead we refer the reader to Ref.\cite{ashtekar/2015} for a study of linear fields on de Sitter space-time.

\subsection{The $f(T^\phi) = 2\lambda (T^\phi)^n$ case with $V_0 \neq 0$}

If the minimum value of the potential $V_0 \neq 0$, and by assuming $m^2 \phi_1 \ll V_0$, the case $f(T^\phi) = 2\lambda (T^\phi)^n$, with $n$ being a constant, does not provide a first order term in $\phi$ in the right hand side of equations (\ref{frt7}) or (\ref{field eq Ricci}). The first two non-null terms of the expansion are the zero and second order terms. Therefore, as in the previous case, there are GWs just with the $+$ and $\times$ polarizations. The overall effect is just a redefinition of $\Lambda$ in the de Sitter background metric as
\begin{equation}\label{Lambda n}
\Lambda = 2 [V_0 - 2\lambda (4V_0)^n].
\end{equation}

\subsection{The $f(T^\phi) = 2\lambda \sqrt{T^\phi}$ case}

Such an $f(T^\phi)$ functionality has already been studied as an $f(R,T)$ gravity case in \cite{shabani/2013}. This particular $f(T^\phi)$ is the only in which the conservation law is respected in a minimal coupling of matter and geometry. This $f(T^\phi)$ exhibits the following equation for $\phi$
\begin{equation}\label{Eq of motion for phi T1/2}
(\sqrt{T^\phi} - 2\lambda) \Box\phi + (\sqrt{T^\phi} - 4\lambda)\frac{\partial V}{\partial \phi} + \lambda \nabla^\mu \phi \nabla_\mu (\ln T^\phi) = 0.
\end{equation} 

Following the same procedure as before, expanding $V(\phi)$ around a non null minimum value $V_0$ and keeping terms up to first order in $\phi$ in the above equation we find
\begin{equation}
\Box \phi + \left(\frac{\sqrt{V_0} - 2\lambda}{\sqrt{V_0} - \lambda}  \right)m^2 (\phi - \phi_0) = 0.
\end{equation}

The solution of this equation is of the form (\ref{model of solution for phi}) with
\begin{equation}
q_\alpha q^\alpha = \left(\frac{\sqrt{V_0} - 2\lambda}{\sqrt{V_0} - \lambda}  \right) m^2,
\end{equation}
and $\lambda \leq \sqrt{V_0}/2$ or $\lambda > \sqrt{V_0}$. However, this solution does note imply any additional polarization states for GWs since   this is a particular case of the precedent section with $n = 1/2$ in equation (\ref{Lambda n}).

On the other hand, if we now adopt $V_0 = 0$ from the beginning we find that $\sqrt{T^\phi}$ is of first order in $\phi$. Then, keeping terms up to first order in $\phi$ in Eq.(\ref{Eq of motion for phi T1/2}) (which is now equivalent to say that $\sqrt{T^\phi} \ll \lambda$) we find
\begin{equation}
\Box \phi + 2m^2(\phi - \phi_0) - \frac{1}{2} \partial^\mu\phi~\partial_\mu \ln(T^\phi) = 0.
\end{equation}

If we assume  a propagating solution like (\ref{model of solution for phi}) for this equation we find that it is identically satisfied only if $m = 0$. However, it does not imply $q_\mu q^\mu = 0$. Accordingly, the curvature scalar is given by
\begin{equation}
R = 12 \lambda \sqrt{q_\mu q^\mu} ~\phi_1 e^{iq_\alpha x^\alpha} + \mathcal{O} (\phi^2)
\end{equation}
and, considering the scalar wave $\phi$ and the GW propagating in the $z$ direction, we choose $q^{\mu} = (\omega, 0, 0, k)$,where $\omega$ is the frequency of the scalar field and $k$ is the $z$ component of the wavevector. Thus, the non-null components of the Ricci tensor are
\begin{equation}
R_{00} = 2\lambda \left(\frac{3\omega^2 - k^2}{\sqrt{\omega^2 - k^2}}\right) \phi_1 e^{iq_\alpha x^\alpha},
\end{equation}
\begin{equation}
R_{33} = 2\lambda \left(\frac{3k^2 - \omega^2}{\sqrt{\omega^2 - k^2}}\right)\phi_1 e^{iq_\alpha x^\alpha},
\end{equation}
\begin{equation}
R_{03} = R_{30} = - 4\lambda \left( \frac{\omega k}{\sqrt{\omega^2 - k^2}} \right) \phi_1 e^{iq_\alpha x^\alpha},
\end{equation}
\begin{equation}
R_{11} = R_{22} = -2 \lambda \sqrt{\omega^2 - k^2}~~ \phi_1 e^{iq_\alpha x^\alpha}.
\end{equation}

Now, by using the above results together with the definitions (\ref{np5})-(\ref{np8}) and with the help of Eqs.(\ref{np9})-(\ref{np13}), we evaluate the NP parameters, namely
\begin{equation}
\Psi_2 = \lambda \sqrt{\omega^2 - k^2} ~\phi_1 e^{iq_\alpha x^\alpha},
\end{equation}
\begin{equation}
\Psi_3 = 0,
\end{equation}
\begin{equation}
\Phi_{22} = -\lambda \left[ \frac{(\omega + k)^2}{\sqrt{\omega^2 - k^2}} \right] \phi_1 e^{iq_\alpha x^\alpha}.
\end{equation}

Note also that since there are no further constrains on the components of the Riemann tensor, $\Psi_4 \neq 0$. Thus, similarly to the $f(R)$ gravity case mentioned above, we are led to conclude that Eq.(\ref{9}) holds once again, but now the presence or absence of the scalar longitudinal mode and of the scalar transversal mode depend on $\lambda$. If $\lambda = 0$ these extra polarization modes disappear and we recover GR theory with the only non-null parameter $\Psi_4$.

It is interesting to notice that in the usual scalar-tensor theories of gravity, the presence of a propagating $\Psi_2$ mode is related to the mass of the scalar field, in such a way that if the mass is not zero, this mode does exist (see \cite{eardley/1973,alves/2010}). On the other hand, we showed that the equations of the $f(R,T^\phi) = -R/4 + 2\lambda \sqrt{T^\phi}$ theory in the linearized regime implied a null mass for the scalar field, but the $\Psi_2$ mode is still present. 

Although there are no initial constrains on $q_\mu q^\mu$, the result we have obtained is not valid for $q_\mu q^\mu = 0$ ($\omega = k$) since the invariant $\Phi_{22}$ diverges. Furthermore, we should have $q_\mu q^\mu = \omega^2 - k^2 > 0$ in order to not violate the causality. Therefore, in the week field regime of this theory, we have a massless scalar field with a speed smaller than the speed of light. As a consequence, since the GW modes associated with $\Psi_2$ and $\Phi_{22}$ have the same speed of the scalar field, they also have a speed $v_{GW} < c$.

\section{GWs in the $f(R,T^\phi) = f_1(R) + f_2(T^\phi)$ theory}

In this section we follow closely the method used in Ref.\cite{alves/2009}. Considering the case for which $f(R,T^\phi) = f_1(R) + f_2(T^\phi)$, the field equations (\ref{frt2}) read
\begin{align}\label{Eq of f(R,Tphi)}
&{f_1}_R R_{\mu\nu} - \frac{1}{2}f_1 g_{\mu\nu} + (g_{\mu\nu} \Box - \nabla_\mu \nabla_\nu){f_1}_R =  \nonumber \\
& -\frac{1}{2} \left[ T^\phi_{\mu\nu} - g_{\mu\nu}f_2(T^\phi) - 2{f_2}_T \nabla_\mu \phi \nabla_\nu \phi \right]
\end{align}
whose corresponding trace is given by
\begin{equation}
{f_1}_R R - 2f_1 + 3 \Box {f_1}_R = -\frac{1}{2} \left(T^\phi - 4 f_2 - 2{f_2}_T\nabla_\alpha \phi \nabla^\alpha\phi \right).
\end{equation}

By restricting $f(R,T)$ to the case $f_1(R) = -  \frac{1}{4} (R - \alpha R^{-\beta})$  and $f_2 (T^\phi) = 2\lambda T^\phi$, and ignoring terms of order two or higher in $\phi$, we find a dynamical equation for the Ricci scalar, namely
\begin{equation}
\Box R^{-(1+\beta)} + \frac{\beta + 2}{3\beta} R^{-\beta} - \frac{1}{3\alpha\beta}R + \frac{8(8\lambda - 1)V_0}{3\alpha\beta} = 0,
\end{equation}
which must be solved in order to verify if there is a propagating GW polarization mode associated with $R$. Considering $\beta \geq 1$, we have $R^{-\beta} \gg R$ in the weak field regime and the above equation now reads (similar calculations can be carried out by assuming other range of values for $\beta$ )
\begin{equation}\label{Eq for psi}
\Box \psi + \frac{\beta + 2}{3\beta} \psi^{\frac{\beta}{1+\beta}} + \frac{8(8\lambda - 1)V_0}{3\alpha\beta} = 0,
\end{equation}
where, for convenience, we have renamed $\psi = R^{-(1+\beta)}$.

Nevertheless, equation (\ref{Eq for psi}) has the following form
\begin{equation}\label{Default Eq for psi}
\Box \psi + \frac{\partial U}{\partial \psi} = 0,
\end{equation}
with the potential given by
\begin{equation}\label{Potential of psi}
U(\psi) = \left[\frac{(\beta + 1)(\beta + 2)}{3\beta(2\beta + 1)}\right] \psi^{\frac{2\beta+1}{\beta + 1}} + \frac{8(8\lambda - 1)V_0}{3\alpha \beta} \psi,
\end{equation}
therefore, since it is Lorentz-invariant, it can be solved by the following method used in Ref.\cite{Rajaraman1982}. Let us first consider the static solution of (\ref{Default Eq for psi})
\begin{equation}
\frac{d^2 \psi}{dz^2} = \frac{\partial U}{\partial \psi},
\end{equation}
which can be written as
\begin{equation}
\frac{1}{2}\left(\frac{d\psi}{dz}\right)^2 = U(\psi).
\end{equation}
Substituting the potential (\ref{Potential of psi}) in the above equation and noticing that the last term of the potential is much smaller than the first, we find that
\begin{align}\label{psi of beta}
 \psi^\frac{1}{2(\beta + 1)} &+ \frac{4(8\lambda - 1)(2\beta + 1)V_0}{\alpha(\beta + 1)(\beta + 2)(2\beta - 1)} \psi^{-\frac{2\beta - 1}{2(\beta + 1)}} = \nonumber \\
& \left[\frac{(\beta + 2)}{6\beta(\beta + 1)(2\beta + 1)}\right]^{\frac{1}{2}}\left(z + C \right),
\end{align}
with $C$ being an integration constant

The most simple solution of the above equation can be found for $\beta = 1$, namely
\begin{equation}\label{Static R}
R(z) = \psi^{-\frac{1}{2}}(z) = \xi^2 \left[ (z + C) \pm \sqrt{(z + C)^2 - 4\sqrt{3}/\xi}\right]^2,
\end{equation}
where
\begin{equation}
\xi = \frac{\alpha}{8\sqrt{3}(8\lambda - 1)V_0}.
\end{equation}
Now, since $R$ is Lorentz invariant, we can obtain a time-dependent solution from the static solution (\ref{Static R}) by considering a Lorentz transformation
\begin{align}
R(t, z) =  \xi^2 \Big\{ & \gamma(z - vt) + C \nonumber \\
&  \pm \sqrt{\left[\gamma(z -vt) + C\right]^2 - 4\sqrt{3}/\xi}~~\Big\}^2,
\end{align}
where $\gamma = (1 - v^2)^{-\frac{1}{2}}$ is the Lorentz factor.

Now, with this result in the field equations (\ref{Eq of f(R,Tphi)}) we are able to find the following relevant components of the Ricci tensor
\begin{equation}
R_{00} = F(t,z)G(t,z) - \frac{1}{2}R(t,z),
\end{equation}
\begin{equation}
R_{33} = v^2F(t,z)G(t,z) + \frac{1}{2}R(t,z),
\end{equation}
\begin{equation}
R_{03} = -vF(t,z)G(t,z),
\end{equation}
where the functions $F(t,z)$ and $G(t,z)$ are given respectively by
\begin{equation}
F(t,z) = \frac{16\gamma^2}{[\gamma(z-vt) + C]^2 - 4\sqrt{3}/\xi}
\end{equation}
and
\begin{equation}
G(t,z) = 1 \pm \frac{1}{4}\frac{[\gamma(z - vt) + C]}{\sqrt{[\gamma(z - vt) + C]^2 - 4\sqrt{3}/\xi}}.
\end{equation}

The components $R_{11}$ and $R_{22}$ are also non-null but  since they do not enter in the calculation of the NP parameters,  we do not quote them here. Finally, from Eqs.(\ref{np5})-(\ref{np8}) and with the help of Eqs.(\ref{np9})-(\ref{np13}), we are able to find the NP parameters
\begin{equation}
\Psi_2 = \frac{1}{12} R(t,z),
\end{equation} 
\begin{equation}
\Psi_3 = 0,
\end{equation}
\begin{equation}
\Phi_{22} = - \frac{1}{4} (1 + v)^2 F(t,z)G(t,z).
\end{equation}

Additionally, since the theory does not exhibit further constrains in the spacetime geometry, we conclude that $\Psi_4 \neq 0$ although it is not possible to obtain its behaviour from the curvature scalar or from the Ricci tensor (because $\Psi_4$ is the NP invariant associated with the Weyl tensor). Therefore, again we find that Eq.(\ref{9}) holds.

Now, the presence or absence of the scalar polarization states do not depend on $\lambda$. This is because they can be generated simply by the particular choice of the function $f_1(R)$ we considered. On the other hand, by taking $\alpha = 0$ we have $R = F = 0$ and then $\Psi_2 = \Phi_{22} = 0$, which is in accordance with our previous assertion that the $f(R,T^\phi) = -R/4 + 2\lambda T^\phi$ theory exhibits only the two usual tensor polarizations of GR. 

It is worth stressing that the value $\beta = 1$ was chosen in Eq.(\ref{psi of beta}) only for simplicity of the subsequent calculations, but it should be remembered that the theory $f(R) = -\frac{1}{4} (R - \alpha R^{-1})$ suffers the well known Dolgov-Kawasaki instability \cite{Dolgov2003}. Although the waveforms for $\Psi_2$ and $\Phi_{22}$ depend on the choice of $\beta$, we do not expect qualitative changes in the final result.

\section{Conclusions}
\label{sec:dis}

With the recent detection of GWs by the Advanced LIGO team \cite{abbott/2016}, a new window to observe the Universe has finally been opened. The high detection rate expected for some events, as the one detected (black hole - black hole merger), allied to some electromagnetic counterparts may lead us to understand Physics at extreme regimes of gravitational fields, densities etc.

The GW spectrum as well as its polarization modes are theory dependent. Previously motivated by shortcomings of standard cosmological scenario, the alternative theories of gravity can also contribute to the study of GWs, being able to generate observables to be corroborated by experiment. 

In this article, we have presented a study of GWs in the $f(R,T)$ and $f(R,T^\phi)$ theories of gravity. The $f(R,T)$ theories consider the gravitational part of the action to be dependent not only on a generic function of $R$, but also on a function of $T$. The dependence on $T$ comes from the consideration of exotic fluids or quantum effects. The $f(R,T)$ models depend on a source term, which represents the variation of the energy-momentum tensor with respect to the metric. On the other hand, in the $f(R,T^\phi)$ theories, it is considered a function of $R$ and of the trace of the energy-momentum tensor of a self-interacting scalar field $\phi$, while the energy-momentum tensor of matter fields enters the field equations in the usual way. In both cases, the field equations of the $f(R)$ gravity are recovered if $T=0$ or $T^\phi = 0$. 

It is the first time that GWs are considered in $f(R,T)$ and $f(R, T^\phi)$ theories. The first steps of this investigation have shown us that, in terms of the polarization modes, it is not possible to distinguish $f(R,T)$ gravity from $f(R)$ gravity. This is because in order to find the number of polarization modes of GWs in a given theory one has to examine the theory in a region far from the source of GWs where $T = 0$. In this regime, $f(R,T)$ gravity retrieves $f(R)$ theory. However, it is expected that it would be possible to distinguish the two theories by analysing the waveforms produced by a given source, a binary system for instance, since the energy-momentum tensor of the source enters in a different manner in the $f(R,T)$ gravity.

In Ref.\cite{ms/2016}, through the introduction of a scalar field, the $f(R,T^\phi)$ theory was considered. In that paper, there is a contribution coming from the $T^\phi$ term, and, consequently, the theory is distinguishable from the $f(R)$ gravity even for $T = 0$ regimes, namely radiation era and vacuum. Starting from such a formulation, we have shown in the present article that indeed it is possible to obtain $f(R,T^\phi)$ gravity information in vacuum regime without necessarily recovering $f(R)$ gravity. By using the field equations of the theory, we have obtained the NP quantities and we have found out extra polarization states of GWs.

As expected, the properties of GWs depend upon the functional form of $f(R,T^\phi)$. In Section 4 we have taken $f(R,T^\phi)=-R/4+f(T^\phi)$ and analysed different forms for $f(T^\phi)$ along with different assumptions for the scalar field potential. For $f(T^\phi)=2\lambda T^\phi$ we showed that the effects of the inclusion of the scalar field up to first order terms are equivalent to consider that the usual CC $\Lambda\rightarrow2(1-8\lambda)V_0$. However, it has already been shown that a CC does not introduce any additional polarization states for GWs (see Ref. \cite{Naf2009}).

It is well known that in a $\Lambda=0$ case, in order to study isolated systems in the weak field regime, one investigates the linearized gravitational fields in Minkowski space-time. Such GW equations are recovered in $f(R,T^\phi)=-R/4+2\lambda T^\phi$ gravity when $\lambda=1/8$ or $V_0=0$. On the other hand, for the $\Lambda>0$ case, it seems natural to replace Minkowski metric with the de Sitter one, as quoted in Section 4.1.

A particular form for $f(T^\phi)$, namely $f(T^\phi) = 2\lambda \sqrt{T^\phi}$ exhibits a quite different scenario. In this theory we have shown that GWs can have two scalar polarization modes (longitudinal and transversal) beyond the usual Einstein polarizations. Nevertheless, it is worth remembering that since $\Psi_2 \neq 0$, the $E(2)$ classification of the theory is ${\rm II}_6$, i.e., $\Psi_2$ is the only observer-independent mode. The presence or absence of all other modes depends on the observer. Additionally, we found out that these scalar polarization modes have a speed $v_{GW} < c$, as in the massive gravity case \cite{depaula/2004}.

Similar results were obtained for the theory $f(R,T^\phi) = -\frac{1}{4}(R - \alpha R^{-\beta}) + 2\lambda T^\phi$, but now the presence or absence of the extra scalar polarization modes do not depend on the presence of the term $2\lambda T^\phi$ since the $f(R)$ gravity also presents these modes. However, the waveforms of the NP parameters $\Psi_2$ and $\Phi_{22}$ we have obtained are quite different from those of the $f(R)$ gravity (as one can compare with the expressions obtained in Ref.\cite{alves/2009}), what could be a way to distinguish between the two theories.

The recent detection of GWs by the LIGO team is consistent with a binary black hole system in general relativity \citep{abbott/2016}. However, because of the similar orientations of the Hanford and Livingston LIGO instruments, the data cannot exclude  the presence of non-Einsteinian polarization modes. In order to determine the polarization content of a signal it is required a network of detectors with different orientations, such as Virgo \cite{Virgo2015}. Also, with only two detectors and in the absence of electromagnetic or neutrino counterpart, there is a large uncertainty in the sky location of the source. As a consequence, there is an uncertainty in the speed of the GWs estimated from the difference of the time of arrival of the signal in each detector, thus $v_{GW} < c$ cannot be ruled out. Therefore, the $f(R,T)$ formalism discussed here, as well as several other alternative theories of gravitation, are not excluded from the point of view of the polarization modes or the speed of GWs. We hope that with the future detection of GWs stronger bounds could be established for such parameters.
\\

{\bf Acknowledgements}
The authors would like to thank the Brazilian agency S\~ao Paulo Research Foundation (FAPESP) for financial support (grant 13/26258-4). PHRSM thanks FAPESP, grant 2015/08476-0, for financial support and O.D. Miranda for suggesting the study of extra polarization states of gravitational waves in $f(R,T)$ gravity. JCNA thanks the Brazilian agency CNPq (308983/2013-0) for partial support.

\appendix
\section{An overview of the Newman-Penrose formalism}
\label{sec:np}


A powerful tool to study the properties of GWs was developed in \cite{eardley/1973}. The basic idea is to analyse all the physically relevant components of the Riemann tensor which cause relative acceleration among test particles. In \cite{eardley/1973}, the authors used a null tetrad basis, which is specially suitable to treat approximately null waves, to calculate the NP quantities \cite{newman/1962,newman/1962b}, which are directly related to the GW polarization states in a given theory. Those quantities are given in terms of the irreducible parts of the Riemann tensor, i.e., the Weyl tensor, the traceless Ricci tensor and the Ricci scalar.

The analysis in \cite{eardley/1973} showed that there are up to six possible modes of polarization for GWs, depending on the theory, which can be corroborated by experiments. Therefore it is possible to categorize a given theory from its non-null NP quantities. 

At a given point, the complex tetrad $({\bf k},{\bf l},{\bf m},{\bf{\bar{m}}})$ is related to the usual cartesian tetrad $({\bf e}_t,{\bf e}_x,{\bf e}_y,{\bf e}_z)$ as

\begin{equation}\label{np1}
{\bf k}=\frac{1}{\sqrt{2}}({\bf e}_t+{\bf e}_z),
\end{equation}
\begin{equation}\label{np2}
{\bf l}=\frac{1}{\sqrt{2}}({\bf e}_t-{\bf e}_z),
\end{equation}
\begin{equation}\label{np3}
{\bf m}=\frac{1}{\sqrt{2}}({\bf e}_x+i{\bf e}_y),
\end{equation}
\begin{equation}\label{np4}
{\bf \bar{m}}=\frac{1}{\sqrt{2}}({\bf e}_x-i{\bf e}_y).
\end{equation}

In general, the NP quantities are independent. In the study of approximately plane waves, there are some differential and symmetrical properties of the Riemann tensor which reduce the number of non-null independent components from $20$ (ten $\Psi$'s, nine $\Phi$'s and $\Lambda$) to six. Therefore we can choose the set $\{\Psi_2,\Psi_3,\Psi_4,\Phi_{22}\}$ to describe, in a given coordinate system, the six independent components of a wave in a given theory. Consequently, in the tetrad basis and in the case of plane waves, the NP quantities are given by

\begin{equation}\label{np5}
\Psi_2=-\frac{1}{6}R_{lklk},
\end{equation}
\begin{equation}\label{np6}
\Psi_3=-\frac{1}{2}R_{lkl\bar{m}},
\end{equation}
\begin{equation}\label{np7}
\Psi_4=-R_{l\bar{m}l\bar{m}},
\end{equation}
\begin{equation}\label{np8}
\Phi_{22}=-R_{lml\bar{m}},
\end{equation} 
with $R_{\alpha\beta\mu\nu}$ being the Riemann tensor. Note that $\Psi_3$ and $\Psi_4$ are complex quantities, so that each of them represents two independent polarization states, one represented by the real part and the other by the imaginary part of $\Psi_3$ and $\Psi_4$. 


Other useful expressions for the NP formalism are the following

\begin{equation}\label{np9}
R_{lk}=R_{lklk},
\end{equation}
\begin{equation}\label{np10}
R_{ll}=2R_{lml\bar{m}},
\end{equation}
\begin{equation}\label{np11}
R_{lm}=R_{lklm},
\end{equation}
\begin{equation}\label{np12}
R_{l\bar{m}}=R_{lkl\bar{m}},
\end{equation}
\begin{equation}\label{np13}
R=-2R_{lk}=-2R_{lklk},
\end{equation}
with $R_{\mu\nu}$ being the Ricci tensor.


\begin{thebibliography}{100}

\bibitem{abbott/2016} B.P. Abbott et al., Phys. Rev. Lett. 116 (2016) 061102.
\bibitem{mm/2014} P.H.R.S. Moraes and O.D. Miranda, MNRAS Lett. 445 (2014) L11.
\bibitem{mm/2014b} P.H.R.S. Moraes and O.D. Miranda, Astrophys. Space. Sci. 354 (2014) 645.
\bibitem{berry/2015} C.P.L. Berry et al., Astrophys. J. 804 (2015) 114.
\bibitem{veitch/2015} J. Veitch et al., Phys. Rev. D 91 (2015) 042003.
\bibitem{cutler/1994} C. Cutler and \'E. E. Flanagan, Phys. Rev. D 49 (1994) 2658.
\bibitem{poisson/1995} E. Poisson and C.M. Will, Phys. Rev. D 52 (1995) 848.
\bibitem{takami/2014} K. Takami et al., Phys. Rev. Lett. 113 (2014) 091104.
\bibitem{lackey/2015} B.D. Lackey and L. Wade, Phys. Rev. D 91 (2015) 043002.
\bibitem{del_pozzo/2013} W. Del Pozzo et al., Phys. Rev. Lett. 111 (2013) 071101.
\bibitem{bauswein/2012} A. Bauswein et al., Phys. Rev. D 86 (2012) 063001.
\bibitem{agathos/2015} M. Agathos et al., Phys. Rev. D 92 (2015) 023012.
\bibitem{read/2009} J.S. Read et al., Phys. Rev. D 79 (2009) 124033.
\bibitem{alves/2009} M.E.S. Alves et al., Phys. Lett. B 679 (2009) 401.
\bibitem{del_pozzo/2011} W. Del Pozzo et al., Phys. Rev. D. 83 (2011) 082002.
\bibitem{konoplya/2016} R. Konoplya and A. Zhidenko, Phys. Lett. B 756 (2016) 350.
\bibitem{yagi/2010} K. Yagi and T. Tanaka, Phys. Rev. D. 81 (2010) 064008.
\bibitem{clifton/2012} T. Clifton et al., Phys. Rep. 513 (2012) 1.
\bibitem{padmanabhan/2003} T. Padmanabhan et al., Phys. Rep. 380 (2003) 235.
\bibitem{kehagias/2004} A. Kehagias, Phys. Lett. B 600 (2004) 133.
\bibitem{arkani-hamed/2000} N. Arkani-Hamed et al., Phys. Lett. B 480 (2000) 193.
\bibitem{harko/2011} T. Harko et al., Phys. Rev. D 84 (2011) 024020.
\bibitem{ms/2016} P.H.R.S. Moraes and J.R.L. Santos, Eur. Phys. J. C 76 (2016) 60.
\bibitem{mc/2016} P.H.R.S. Moraes and R.A.C. Correa, Astrophys. Space. Sci. 361 (2016) 91.
\bibitem{moraes/2016} P.H.R.S. Moraes, Int. J. Theor. Phys. 55 (2016) 1307.
\bibitem{moraes/2015} P.H.R.S. Moraes, Eur. Phys. J. C 75 (2015) 168.
\bibitem{moraes/2014} P.H.R.S. Moraes, Astrophys. Space. Sci. 352 (2014) 273.
\bibitem{shamir/2015} M.F. Shamir, Eur. Phys. J. C 75 (2015) 354.
\bibitem{baffou/2015} E.H. Baffou et al., Astrophys. Space. Sci. 356 (2015) 173.
\bibitem{singh/2014} C.P. Singh and P. Kumar, Eur. Phys. J. C 74 (2014) 3070.
\bibitem{shabani/2014} H. Shabani and M. Farhoudi, Phys. Rev. D 90 (2014) 044021.
\bibitem{eardley/1973} D.M. Eardley et al., Phys. Rev. D 8 (1973) 3308.
\bibitem{newman/1962} E. Newman and R. Penrose, J. Math. Phys. 3 (1962) 566.
\bibitem{newman/1962b} E. Newman and R. Penrose, J. Math. Phys. 4 (1962) 998.
\bibitem{alves/2010} M.E.S. Alves et al., Class. Quant. Grav. 27 (2010) 145010.
\bibitem{depaula/2004} W.L.S. de Paula et al., Class. Quant. Grav. 21 (2004) 4595.
\bibitem{nojiri/2009} S. Nojiri et al., Phys. Lett. B 681 (2009) 74.
\bibitem{nojiri/2007} S. Nojiri and S.D. Odintsov, Phys. Lett. B 657 (2007) 238.
\bibitem{harko/2014} T. Harko, Phys. Rev. D 90 (2014) 044067.
\bibitem{Naf2009} J. N\"af, P. Jetzer, M. Sereno, Phys. Rev. D 79 (2009) 024014.
\bibitem{ashtekar/2015} A. Ashtekar et al., Phys. Rev. D 92 (2015) 044011.
\bibitem{shabani/2013} H. Shabani and M. Farhoudi, Phys. Rev. D 88 (2013) 044048.
\bibitem{Rajaraman1982} R. Rajaraman, Solitons and Instantons: An Introduction to Solitons and Instantons in Quantum Field Theory, Elsevier Science Publishers, Amsterdam, The Netherlands (1982)
\bibitem{Dolgov2003} A.D. Dolgov and M. Kawasaki, Phys. Lett. B 573 (2003) 1-4.
\bibitem{Virgo2015} F. Acernese et al. (Virgo), Class. Quant. Grav. 32 (2015) 024001.


\end{thebibliography}
\end{document}